\documentclass[manuscript,screen]{acmart}
\AtBeginDocument{%
  }

\setcopyright{acmlicensed}
\copyrightyear{2018}
\acmYear{2018}
\acmDOI{XXXXXXX.XXXXXXX}
\acmConference[Conference acronym 'XX]{Make sure to enter the correct
  conference title from your rights confirmation email}{June 03--05,
  2018}{Woodstock, NY}
\acmISBN{978-1-4503-XXXX-X/2018/06}
\usepackage{listings}
\definecolor{keywordcolor}{RGB}{127,0,85}  
\definecolor{commentcolor}{RGB}{63,127,95} 
\definecolor{stringcolor}{RGB}{42,0,255}  
\usepackage{booktabs}
\usepackage{tabularx}
\usepackage{multirow}
\usepackage{hyperref}  
\usepackage{float}
\usepackage{floatpag}
\usepackage{comment}
\usepackage{listings}
\usepackage{tabularx}
\usepackage{tcolorbox}
\usepackage{siunitx}
\usepackage{algorithmic}
\usepackage{graphicx}
\usepackage{textcomp}
\usepackage{xcolor}
\usepackage{comment}

\lstdefinelanguage{Solidity}{
    keywords={pragma, solidity, contract, function, address, uint, require, external, public, private, internal, returns, event, emit, onlymanyowners, keccak256, msg, data, balance, transfer},
    keywordstyle=\color{keywordcolor}\bfseries,
    comment=[l]{//},
    morecomment=[s]{/*}{*/},
    commentstyle=\color{commentcolor}\itshape,
    stringstyle=\color{stringcolor},
    morestring=[b]",
    morestring=[b]',
    basicstyle=\ttfamily\small,
    numbers=left,
    numberstyle=\tiny,
    stepnumber=1,
    breaklines=true,
    frame=single,
    tabsize=2
}

\definecolor{lightblue}{rgb}{0.9,0.95,1}
\definecolor{lightgray}{rgb}{0.95,0.95,0.95}
\definecolor{highlight}{rgb}{0.8,0.0,0.0}

\lstdefinestyle{textonly}{
  basicstyle=\ttfamily\small,
  breaklines=true,
  showstringspaces=false,
  frame=single,
  moredelim=[is][\color{highlight}]{\{\{}{\}\}},
}
\usepackage{xcolor}

\begin{document}


%\title{The Name of the Title Is Hope}
\title{Beyond Code Similarity: Benchmarking the Plausibility, Efficiency, and Complexity of LLM-Generated Smart Contracts}

\author{Francesco Salzano}
\email{francesco.salzano@unimol.it}
\authornote{Corresponding author.}
\orcid{0000-0002-1029-4861}
\affiliation{%
  \institution{University of Molise}
  \city{Pesche}
  \state{Isernia}
  \country{Italy}
}

\author{Simone Scalabrino}
\email{simone.scalabrino@unimol.it}
\orcid{0000-0003-1764-9685}
\affiliation{%
  \institution{University of Molise}
  \city{Pesche}
  \state{Isernia}
  \country{Italy}
}

\author{Rocco Oliveto}
\email{rocco.oliveto@unimol.it}
\orcid{0000-0002-7995-8582}
\affiliation{%
  \institution{University of Molise}
  \city{Pesche}
  \state{Isernia}
  \country{Italy}
}

\author{Remo Pareschi}
\email{remo.pareschi@unimol.it}
\orcid{0000-0002-4912-582X}
\affiliation{%
  \institution{University of Molise}
  \city{Pesche}
  \state{Isernia}
  \country{Italy}
}

\renewcommand{\shortauthors}{Salzano et al.}

\begin{abstract}

Smart Contracts are critical components of blockchain ecosystems, with Solidity as the dominant programming language. While LLMs excel at general-purpose code generation, the unique constraints of Smart Contracts, such as gas consumption, security, and determinism, raise open questions about the reliability of LLM-generated Solidity code. Existing studies lack a comprehensive evaluation of these critical functional and non-functional properties.

We benchmark four state-of-the-art models under zero-shot and retrieval-augmented generation settings across 500 real-world functions. Our multi-faceted assessment employs code similarity metrics, semantic embeddings, automated test execution, gas profiling, and cognitive and cyclomatic complexity analysis.

Results show that while LLMs produce code with high semantic similarity to real contracts, their functional correctness is low: only 20\%–26\% of zero-shot generations behave identically to ground-truth implementations under testing. The generated code is consistently simpler, with significantly lower complexity and gas consumption, often due to omitted validation logic. Retrieval-Augmented Generation markedly improves performance, boosting functional correctness by up to 45\% and yielding more concise and efficient code.

Our findings reveal a significant gap between semantic similarity and functional plausibility in LLM-generated Smart Contracts. We conclude that while RAG is a powerful enhancer, achieving robust, production-ready code generation remains a substantial challenge, necessitating careful expert validation.
\end{abstract}




\received{20 February 2007}
\received[revised]{12 March 2009}
\received[accepted]{5 June 2009}

\maketitle

\section{Introduction}
Since its birth with Bitcoin~\cite{nakamoto2008bitcoin}, blockchain technology has emerged as a disruptive innovation in numerous domains, including Decentralized Finance and data certification, handling high-stakes assets like cryptocurrencies and sensitive data~\cite{barboni2022smart}. Smart Contracts (SCs) are central to this ecosystem. SCs are Turing-complete programs executed on blockchains, in which consensus protocols guarantee valid execution through validated transactions~\cite{chen2020defining}. SCs are replicated identically across decentralized network nodes and are designed to automatically trigger predefined actions in response to specific events~\cite{zou2019smart}. Interactions with SCs occur via transactions, validated by blockchain nodes, with outcomes recorded on each node's copy of the distributed ledger. Solidity is the most widely used language for SC development~\cite{wohrer2020domain}. Differently from traditional programs, SCs might be directly executed by the end users. Therefore, Solidity provides a special tag, \texttt{@notice} tag -- part of its NatSpec format\footnote{https://docs.soliditylang.org/en/latest/natspec-format.html} -- through which developers can provide non-technical documentation of the function behavior, guiding end-users before contract execution~\cite{hu2021automating}. 

Recent years have also seen rapid advancements in Large Language Models (LLMs), which have shown remarkable capabilities in code-related tasks. Tools like GitHub Copilot and ChatGPT have demonstrated the ability to generate code valid snippets~\cite{corso2024generating,mastropaolo2023robustness}, assist in software comprehension~\cite{nam2024using}, and even detect vulnerabilities across various programming languages~\cite{chen2025chatgpt,lu2024grace}.

The potential of LLMs in generating general-purpose programs (e.g., in Java) has already been widely evaluated~\cite{corso2024generating}. On the other hand, the literature provides only limited evidence of the capability of such approaches to automatically generate SCs. SC development introduces unique challenges: Syntactic and semantic plausibility are not enough, and security, gas-efficiency, and robustness against adversarial inputs are of primary importance to grant the success of a SC. Recent work has begun exploring these issues, evaluating LLMs for tasks ranging from vulnerability detection~\cite{chen2025chatgpt} to automatic comment generation~\cite{zhao2024automatic} and contract generation~\cite{napoli2024leveraging}.
Given the syntactic similarity between Solidity and other general-purpose programming languages such as JavaScript, it is theoretically expected that LLMs have the potential to succeed in this task. However, it is completely unclear how the generated programs perform in terms of gas-consumption -- as mentioned, a critical property for SCs -- and maintainability.

In this study, we aim to fill this gap by systematically evaluating the capabilities of a set of LLMs in generating Solidity SCs from natural language descriptions. Our objective is not only to understand to what extent correct SCs can be generated, but also to assess how gas-efficient and maintainable such generated programs are.
To this aim, we test four state-of-the-art LLMs, covering both closed- and open-source ones, namely ChatGPT-4o, Gemini 1.5 Flash, CodeLlama, and DeepSeek-Coder-v2. We experiment LLMs in two modalities: (i) we use a simple zero-shot prompt in which we plainly ask those models to generate a Solidity with a given \textit{notice}; (ii) we adopt retrieval-augmented generation (RAG) and provide the model with functionally similar SCs besides the previously-mentioned documentation.

First, we measure how closely the generated functions resemble existing implementations across multiple dimensions. \textbf{Code similarity} is assessed via BLEU, structural similarity using Tree Edit Distance (TED) on abstract syntax trees (ASTs), and semantic similarity through SmartEmbed embeddings~\cite{gao2019smartembed}. We also assess the \textbf{code plausibility} by running automatically generated tests, to understand whether, in spite of differences, the generated SCs are \textit{plausible} solutions. We refer to plausiblity since generated contracts
require further analysis to ensure that they are fully correct, i.e., if these are acceptable to developers~\cite{liu2020efficiency}.
Then, we focus on non-functional properties. We measure the \textbf{gas-efficiency} to understand whether the generated SCs are practically viable solutions and \textbf{code complexity} -- evaluated with cyclomatic and cognitive complexity metrics -- to estimate their maintainability.

Our results show that, although all evaluated LLMs can produce Solidity functions highly similar to real contracts, their functional plausibility remains limited in the absence of additional context.
In a zero-shot setting, LLMs generate \textit{plausible} SCs (i.e., for which no test fails) up to 9.09\% of the cases with up to 25.53\% of correct function' call output. When using RAG, these percentages reach up to 45.19\% and 23.08\% with DeepSeek-coder-v2. The gas consumption is substantially lower than the one of manually-written SCs. 
We observed that RAG substantially improves plausibility rates and reduces gas consumption, narrowing the gap between generated and reference implementations. 

Overall, our study contributes a systematic benchmark of LLM performance on SC generation, considering similarity, complexity, and plausibility, and provides evidence that retrieval-based prompting is a promising direction for improving practical viability.
All data, code, and artifacts involved in our work are available in our \textbf{replication package}~\cite{replication_package}.

%%%%%%%%%%%

\section{Related Work}
This section provides an overview of previous studies and research that are relevant to our current study, highlighting their key findings and methodologies that have influenced or informed our work.
As software engineering is living in the LLM era, multiple contributions to the field regarding LLM evaluations.

Indeed, LLMs have been demonstrated to be useful in several code-related tasks. For instance, Corso et al. conducted a comprehensive empirical study comparing four state-of-the-art AI-based code assistants, GitHub Copilot, Tabnine, ChatGPT, and Google Bard, on the task of generating Java methods from real-world open-source projects~\cite{corso2024generating}. Their methodology involved constructing a curated dataset of 100 Java methods with varying complexity and dependency contexts. %Such research assessed code generation quality along multiple dimensions, including plausibility, efficiency, complexity, size, and stylistic similarity to developer-written code. 
The results showed that while Copilot was the most effective overall, each assistant succeeded in cases where others failed. However, performance dropped notably for methods with external dependencies.
Mastropaolo et al. investigated the robustness of GitHub Copilot concerning changes in natural language input~\cite{mastropaolo2023robustness}. They found that semantically equivalent paraphrases of method descriptions lead to different generated code in 46\% of cases, with potential loss of correct predictions in about 28\%.
Fakhour et al. introduced TiCoder, an interactive test-driven workflow to improve the accuracy of LLM-generated code~\cite{fakhoury2024llm}. Their approach helps clarify user intent by generating and validating tests that guide LLMs in ranking and pruning code suggestions. Through a user study and a large-scale empirical evaluation, they demonstrated that TiCoder significantly increases code plausibility and reduces cognitive load, achieving up to 45.7\% improvement in \textbf{pass@1} accuracy across several models and datasets. Such a study reinforces the importance of an interactive mechanism when dealing with AI-assisted coding.
%%%%%% LLM AND SMART CONTRACTS
In the field of SCs, LLMs are mainly evaluated in security vulnerability detection. For example, Chen et al. evaluated different versions of ChatGPT, including 3.5, 4, and 4o~\cite{chen2025chatgpt}, using the SmartBugs Curated dataset~\cite{durieux2020empirical}. This dataset contains 142 SCs manually labeled with vulnerability annotations. The results showed that ChatGPT achieved high recall but relatively low precision. 

Furthermore, Zhao et al. proposed \textit{SCCLLM}, an approach dedicated to automatically generating comments for Solidity SCs~\cite{zhao2024automatic}, which is based on ChatGPT-3.5-turbo and in-context learning, utilizing the top 5 code snippets retrieved from a historical corpus by considering semantic, lexical, and syntactical features. The effectiveness of \textit{SCLLM} was assessed with human studies alongside automatic performance metrics.
%code summarization 
Hu et al. proposed SMARTDOC, a Transformer-based approach enhanced with a pointer mechanism and transfer learning to automatically generate user notices (end-user-facing natural language descriptions) for SC functions in Solidity~\cite{hu2021automating}.

Regarding SC code generation, Napoli et al. presented a pipeline that utilizes ChatGPT, aiming to make Solidity development accessible to those unfamiliar with it~\cite{napoli2024leveraging}. Such a pipeline demonstrated to be capable of producing 98\% of compilable SCs without high-impact vulnerabilities. Moreover, in this study, variations in the value of the temperature were shown to have negligible effects on the generations.

%%% automated update

Nevertheless, all these contributions deeply investigated the capabilities of LLMs in SC code-related tasks. Still, comprehensive empirical evaluations of generation ability are missing, and no study has delved into the functional plausibility of Solidity LLM-generated code.

\section{Study Design}

The \textit{goal} of this research is to evaluate LLMs for SC code generation in terms of (i) \textit{plausibility}, (ii) \textit{efficiency}, and (iii) \textit{maintainability}.

The following research questions guide our work:
\begin{itemize}
    \item \textbf{RQ$_1$: } How similar are the generated functions to the ones extracted from real-world contracts?
     \item \textbf{RQ$_2$: } To what extent is the generated code syntactically and functionally correct?
     \item \textbf{RQ$_3$: } What is the gas consumption of the generated smart functions?
    \item \textbf{RQ$_4$: } What is the Cognitive and Cyclomatic complexity of the generated Smart Contract functions?
    %\item \textbf{RQ$_5$: } How does the effectiveness of the generation vary when using RAG to generate code?
\end{itemize}

\subsection{Study Context}

The context of our research consists of two types of experimental \textit{objects}:  
(i) Large Language Models for code generation, and  
(ii) a dataset of pairs of Smart Contract (SC) notices and corresponding Solidity functions.

As for the LLMs, we included both open and closed models. For closed models, we used ChatGPT-4o and Gemini 1.5 Flash, for which architectural details such as parameter count are not publicly disclosed. For open models, we selected CodeLlama (13B parameters) and DeepSeek-Coder-v2 (16B parameters). Our selection criterion was to cover a representative spectrum of model types: open versus closed, code-specialized versus general-purpose, and, within each category, both earlier and more recent model generations.

Regarding the dataset, we consider pairs of notices and functions extracted from the dataset presented by Hu et al.~\cite{hu2021automating}, which comprises 7,878 instances. Although released in 2021, it remains highly relevant, as our investigation also targets the impact of input formulation on Solidity code generation. In our setting, the baseline consists of generating code from notices, which convey the semantic intent of functions~\cite{hu2021automating}.

In the Hu et al. dataset, notices and functions are stored in separate files and share index-based correspondence. We merged them into unified notice–function pairs and applied filtering criteria prior to sampling. Specifically, we removed instances in which the function was a modifier or the notice contained fewer than eight words. This ensured that the dataset comprised complete Solidity functions and excluded overly simple cases, such as getters, which typically have minimal semantic content. To determine the word-count threshold, one author manually inspected a stratified sample of 100 instances and found that notices containing fewer than eight words were overly simple or lacked sufficient semantic content in approximately 66\% of cases.

Out of the resulting 4,509 valid instances, we randomly extracted a statistically significant sample of 500 examples. We selected a sample size that ensures a margin of error below 5\% at a 95\% confidence level; specifically, a sample of 500 instances yields an estimated margin of error of 4.38\%. Table~\ref{tab:complexity_stats} reports descriptive statistics for the complexity of the resulting stratified sample. Notably, all functions in this sample originate from real-world projects and correspond to code that is currently deployed on the blockchain.

\begin{table}[h]
\centering
\caption{Descriptive statistics for complexity and code size metrics.}
\label{tab:complexity_stats}
\begin{tabular*}{\columnwidth}{@{\extracolsep{\fill}}lrrrrr}
\toprule
\textbf{Metric} & \textbf{Mean} & \textbf{Mode} & \textbf{Std. Dev.} & \textbf{Min} & \textbf{Max} \\
\midrule
Cyclomatic Complexity & 3.076 & 1.0 & 2.5169 & 1.0 & 18.0 \\
Cognitive Complexity (max) & 2.004 & 0.0 & 2.7748 & 0.0 & 29.0 \\
LOC (without spaces and comments) & 7.752 & 3.0 & 5.6606 & 1.0 & 50.0 \\
\bottomrule
\end{tabular*}
\end{table}

\begin{figure}
\begin{lstlisting}[language=Solidity]
function sendEther(address to, uint value) 
    external 
    onlymanyowners(keccak256(msg.data)) 
{
    require(address(0) != to);
    require(value > 0 && this.balance >= value);
    to.transfer(value);
    EtherSent(to, value);
}
\end{lstlisting}
\caption{Example of a function whose syntax is disrupted in the original dataset.}
\label{lst:sendEther}
\end{figure}

In the original dataset, functions are processed in a way that disrupts correct Solidity syntax and formatting, as illustrated in Figure~\ref{lst:sendEther}. To restore syntactically valid Solidity code (e.g., replacing constructs such as \texttt{this . balance} with \texttt{this.balance}), we employed a self-hosted CodeLlama-14B model using a one-shot learning setup. We manually formatted one representative function to create a transformation example, then used the model to automatically repair the remaining instances. All generated fixes were subsequently validated through manual inspection to ensure dataset reliability.

\subsection{Experimental Procedure}
% This section provides an overview of the methodology used to conduct the experiments.
% \subsubsection{Interacting with the LLMs and Generating the code}

For each LLM, we instruct the model to refine and optimize the generation prompt, as proposed by Chen et al.~\cite{chen2025chatgpt}. All interactions with the LLMs are conducted via APIs, including those deployed locally. We set the temperature of all models to 0 because it increases the reproducibility of the output. Besides, Chen et al.~\cite{chen2025chatgpt} recently showed that such a setting improves ChatGPT's performance in SC vulnerability detection~\cite{chen2025chatgpt}.
The requests to Gemini and ChatGPT are sent to the official endpoints. The other LLMs are deployed using Ollama\footnote{\url{https://github.com/ollama/ollama}}.
The local-deployed LLM (CodeLLama and Deepseek-Coder-v2) generation tasks were performed on a workstation equipped with a Ryzen 9-9900x, 32GB RAM, and NVIDIA 4070 GPU.

We adopt two code generation settings:
\begin{enumerate}
    \item \textbf{Baseline (Zero-shot):} A plain prompt with only the system message and the \texttt{@notice} tag, used to assess the standalone generation ability of each LLM.
    \item \textbf{Retrieval-Augmented Generation:} We enrich the prompt by retrieving semantically similar examples SCs from an extraneous dataset (the one by Liu et al. \cite{liu2021smart}).
\end{enumerate}
The general structure of the baseline prompt is shown in Figure~\ref{lst:generation-prompt}.

\begin{figure}
\begin{lstlisting}[style=textonly, backgroundcolor=\color{lightgray}]
"{{system}}": "You are a Solidity expert and your task is to write secure, gas-efficient, and well-documented smart contract code in the Solidity language."
"{{prompt}}": "Please implement the following Solidity function: {{<function>}}. Provide the full function implementation only, without explanations or comments."
\end{lstlisting}
\caption{Prompt structure used for baseline code generation.}
\label{lst:generation-prompt}
\end{figure}
\unskip To support and potentially improve the generation process, we designed a RAG pipeline. Starting from a ground truth \texttt{@notice}, we retrieve the top-$5$ most semantically similar notices from a large corpus of real-world functions to mitigate impacts of potentially incoherent comments in the domain-specific-dataset, using \texttt{all-MiniLM-L6-v2} embeddings and cosine similarity. If fewer than 5 relevant matches are found in \texttt{@notice} fields, we fallback to full comment search. We also ensure diversity by removing duplicates and exact matches (similarity = 1.0).
The retrieval corpus is derived from the dataset of Liu et al.~\cite{liu2021smart}, containing over 40k real-world SCs. We extract function definitions and associated comments to build the retrieval context.
The retrieved examples, comments, and original notice are injected into the prompt template shown in Figure~\ref{lst:rag-generation-prompt}, reused for all LLMs.

\begin{figure}
\begin{lstlisting}[style=textonly, backgroundcolor=\color{lightgray}]
Generate a Solidity smart contract function based on the following summary, example code snippets, and associated comments.

=== Code Summary ===
{{code_summary}}
=== Example Code (from retrieved context) ===
{{retrieved_code_examples}}
=== Developer Comments ===
{{retrieved_comments}}
=== Original Notice ===
{{original_notice}}

Follow Solidity best practices, use modifiers and roles as needed, and comply with version ^0.8.0.
\end{lstlisting}
\caption{Prompt structure used for generation with RAG.}
\label{lst:rag-generation-prompt}
\end{figure}

\subsubsection{RQ$_1$: Code Similarity}
\label{subsec:rq1}
To assess the code similarity between the generated functions and those representing our ground truth, we leverage three metrics: BLEU, Semantic Similarity, and TED.
BLEU is a popular metric for evaluating the quality of AI text by comparing it to reference code content and a candidate-generated code using n-gram precision. To obtain BLEU, the \textit{nltk} Python library was employed.  We are aware that CodeBLEU would be better; unfortunately, there is no CodeBLEU metric tailored to Solidity to date.
To evaluate the semantic closeness between the generated and reference code functions, we compute the \textit{Semantic Similarity}, a metric that quantifies how similarly two code snippets express meaning, regardless of lexical or syntactic differences.
For this purpose, we use \textit{SmartEmbed}~\cite{gao2019smartembed}, a semantic embedding system tailored for Solidity code, to generate vector representations of both the ground truth and the generated functions. Each code snippet is processed through \textit{SmartEmbed}, which outputs high-dimensional embeddings that capture their semantic content. We then compute the semantic similarity between these embeddings using a normalized Euclidean distance, as it is directly included in SmartEmbed.

TED is a metric that measures how different two tree structures are by computing the minimum number of operations required to transform one tree into another, widely used in ASTs. 
To compute TED, we leveraged \textit{zss}, a Python library that computes TED, using the \textit{Zhang-Shasha} algorithm~\cite{zhang1989simple}. We set the cost of substitution fixed to 1, independently of the type of node.
% To reach the calculation of TED, ASTs are needed. To obtain them, solidity\_parser\footnote{https://pypi.org/project/solidity-parser/} is used to parse the source code into an AST.

LLMs often divide the generated solution into more than one function, even though we explicitly asked them to generate only one. Specifically, we noticed that they often generate an internal function with the core implementation and then a main function that performs additional checks and calls the internal function. In this context, we were more interested in the internal function of the core logic of the program. To find it when more than one function is returned, we use a heuristic that selects the function with the highest number of statements. We manually checked 50 examples to confirm the effectiveness of such a heuristic.

To answer RQ$_1$, we report the descriptive statistics (mean, minimum, maximum, and standard deviation) of BLEU, TED, and SmartEmbed for each LLM we considered. To assess the impact of contextual information on similarity metrics, we repeat the above procedure by applying an RAG strategy. In this setting, the same evaluation metrics are computed on the functions generated with RAG. This allows for a direct comparison between the baseline (zero-shot) and RAG-based generation across all LLMs, enabling us to quantify the benefits of retrieved context in improving structural and semantic alignment with ground truth code.

\subsubsection{RQ$_2$: Code plausibility}
\label{subsubsec: correctness}

To systematically evaluate the code plausibility of the generated SCs, we designed an automated pipeline that compiles, deploys, and executes the contracts. As in RQ$_1$, we repeat the entire procedure using RAG-enhanced prompts to compare plausibility between the baseline and RAG settings, enabling a direct evaluation of the functional benefits introduced by contextual retrieval. Specifically, for every generated contract, we dynamically compile it, deploy the artifact to a local Ethereum-like blockchain, and run tests aimed at checking behavioral equivalence between the generated contracts and their corresponding ground-truth instances.

We first identify which functions in the ground-truth dataset are compilable. Since individual functions cannot be compiled in isolation, we wrap them in minimal contracts and reconstruct the necessary compilation units. This requires declaring all state variables (e.g., roles, counters, ownership models) and defining appropriate initialization logic within the constructor. To achieve this, we adopt the \texttt{Qwen-2.5-Coder} model, which is instructed to: (i) add or rewrite constructors, (ii) correctly invoke parent constructors when needed (e.g., \texttt{Ownable()} or \texttt{ERC20("MyToken", "MTK")}), (iii) initialize all state variables with non-corner-case defaults, (iv) avoid altering the logic of non-constructor code, and (v) ensure deployability without runtime exceptions. The full prompt used for constructor injection is shown in Figure~\ref{lst:constructor-injection-prompt}.

\begin{figure}
\begin{lstlisting}[style=textonly, backgroundcolor=\color{lightgray}]
You are a Solidity expert. Your task is to analyze the provided Solidity contract and do the following:
If the contract does not contain a constructor, add one.
If the contract already contains a constructor, rewrite it.
In the constructor, initialize all instance (state) variables to fixed, safe, non-corner-case values.
If the contract inherits from any parent contracts such as Ownable, Pausable, ERC20, or others, you must properly invoke their constructors using the correct syntax (e.g., Ownable(), ERC20('TokenName', 'TKN')).
Initialization rules:
uint/uint256: Set to 1 (never 0)
address: Use these fixed values in order: 0x111..., 0x222..., 0x333...
bool: Set to true
string: Set to 'initialized'
bytes32: Set to bytes32('init')
Avoid default zero values
Do not touch imports and remove 'abstract' if present
Return only the modified contract code without markdown
Contract:
{{contract_code}}
\end{lstlisting}
\caption{Prompt used for constructor injection.}
\label{lst:constructor-injection-prompt}
\end{figure}

\begin{figure}
\begin{minipage}{\linewidth}
\begin{lstlisting}[language=Solidity]
address permissionManager = new GeneralPermissionManager{salt: bytes32(uint256(data))}(msg.sender, address(polyToken));
\end{lstlisting}
\end{minipage}
\caption{Dependency declaration example.}
\label{fig:example:compilation}
\end{figure}

The purpose of the accompanying manual inspection was not to assess the correctness of the generated constructors, but to evaluate the robustness of the prompting strategy in producing constructor code that preserves the original function logic. We followed an iterative refinement process: we drafted an initial prompt, generated constructors, manually reviewed a sample of 50 outputs, and adjusted the prompt to eliminate recurring error patterns. This cycle was repeated four times until no further issues emerged. After generation, we additionally ensured consistent initialization of fund-related variables, setting user balances to 1,000 and the \texttt{totalSupply} to $10^{18}$ to prevent systematic reverts during execution.

At the end of this process, we compiled all artificially reconstructed contracts—both for the generated functions and the ground-truth ones—and obtained 198 valid compilable instances. Compilation errors were primarily due to missing project dependencies, such as references to external contracts (Figure~\ref{fig:example:compilation}), as well as incorrect \texttt{struct} declarations and invalid \texttt{internal} or \texttt{super()} invocations.

\vspace{1ex}

We deploy each contract to a local Ethereum-like testing environment using \texttt{Ganache}. The \texttt{pragma} directive is automatically parsed to identify the required Solidity compiler version, which is installed via the \texttt{solcx} Python package. Compilation is followed by deployment and transaction execution using the Web3.py interface. All experiments were conducted on a machine with an Intel i5 processor and 16 GB of RAM. Unless otherwise specified, tests are executed using the 0.8.x family of the Solidity compiler to align with the evolving language standard.

To compare behavioral differences between generated and reference contracts, we construct a type-aware set of test inputs. For each function, we generate two corner-case inputs: one assigning minimum values and one assigning maximum values to all numeric parameters. For parameters of type \texttt{address}, we use \texttt{0} as the sole corner case; consequently, functions consisting exclusively of address parameters yield a single corner-case input. We then sample 10 additional inputs with random values drawn from the respective parameter domains. We do not evaluate the adequacy of the resulting test cases, as such an assessment would require specialized frameworks and systematic criteria (e.g., coverage metrics), which fall outside the scope of this study and would drastically reduce scalability.

For numeric parameters (e.g., \texttt{uint256}), example values include \texttt{0}, \texttt{1}, and $2^{255}$; \texttt{string} parameters are drawn from a predefined corpus and random ASCII sequences, while \texttt{address} values are converted to valid checksum addresses. Test inputs are generated independently for each parameter and combined to produce a set of 10 test cases per function. For every test, we execute the function on both the generated and reference versions and record return values as well as any exceptions thrown. Functional behavior is considered preserved for a given input if both contracts return identical outputs (or revert in the same manner). A generated function is considered \emph{plausibly equivalent} if it matches the ground truth on all generated tests.

Although the number of test cases per function is relatively small, each execution incurs non-negligible overhead: before each test batch, we initialize a fresh Ganache instance and allow a 5-second stabilization period to minimize nondeterministic noise (e.g., timing-dependent behavior).

To answer RQ$_2$, we report, for each LLM, the percentage of generated functions that exhibit full behavioral plausibility over our test suite. Our objective is not to provide a formal equivalence guarantee—no finite set of tests could ensure such a result—but to detect clear behavioral mismatches in generated contracts. Despite its limitations, this testing methodology represents a meaningful advancement over prior work, which entirely omitted behavioral validation.

\subsubsection{RQ$_3$: Gas Consumption }

To systematically evaluate the runtime cost of SC execution, we start from the Solidity compilation units defined to answer RQ$_2$. Again, we deploy each contract to the local Ethereum-like environment powered by \texttt{Ganache}, as already done by Di Sorbo et. al.~\cite{di2022profiling}.
Using the Web3 interface, we enumerate all public and external methods declared in the contract ABI.
For each function, we use the same test cases adopted to check plausibility in RQ$_2$. We obtain, for each test case, the gas consumption via the \texttt{estimateGas} interface. To answer RQ$_3$, we report summary statistics (mean, minimum, maximum) regarding the consumed gas. We also report the gas consumed by the reference functions to make them comparable. We report the results both for \textit{all} the generated functions and only for the functions that only for the functions that are correct according to the results of RQ$_2$. This second value is, theoretically, more reliable since it only focuses on functions for which the behavior is plausibly the same, thus any difference might not be explained by the fact that the two functions have different features.
To complement the baseline analysis, we also measure gas consumption for functions generated with RAG. This comparison serves to highlight whether the additional context leads to more efficient or costlier executions.

\subsubsection{RQ$_4$: Complexity}

To answer RQ$_4$, we evaluate the structural complexity of the generated functions using two complementary measures: \textit{Cyclomatic Complexity} and \textit{Cognitive Complexity}. Cyclomatic Complexity quantifies the number of independent execution paths in a function's control-flow graph, serving as a standard indicator of structural complexity and maintainability~\cite{gill1991cyclomatic}. In our implementation, the Cyclomatic Complexity score is incremented whenever specific flow-altering constructs are encountered\footnote{\texttt{if}, \texttt{for}, \texttt{while}, \texttt{do}, \texttt{case}, \texttt{require}, \texttt{assert}, \texttt{\&\&}, \texttt{||}, \texttt{catch}, \texttt{try}}.  
Commonly used libraries could not be employed directly, as they do not capture Solidity-specific constructs—most notably \texttt{require}—that introduce implicit branching.

Cognitive Complexity, originally proposed by SonarSource~\cite{sonarsource2017cognitive}, aims to measure code understandability by accounting for nesting depth, control-flow structure, and readability considerations. Unlike Cyclomatic Complexity, which counts independent paths, Cognitive Complexity penalizes nested control structures and other patterns known to affect human comprehension. Prior studies have shown it to be a strong proxy for human-perceived understandability~\cite{munoz2020empirical}. We therefore use Cyclomatic Complexity as a conventional structural baseline, also adopted by Corso et al.~\cite{corso2024generating}, and Cognitive Complexity as a complementary measure that more closely reflects human reasoning.

Although Cognitive Complexity was designed for modern programming languages, its formulation does not directly incorporate Solidity-specific branching constructs. To address this limitation, we re-implemented the metric following its original design~\cite{sonarsource2017cognitive}, extending it to include Solidity flow-breaking elements such as \texttt{require} and \texttt{assert}. Our implementation increases complexity scores for each disruption of the linear flow and applies additional penalties based on nesting level, ensuring semantic alignment with the original intent of the metric.

We chose not to rely on Slither’s complexity metrics~\cite{feist2019slither} because they require fully compilable contracts, whereas our complexity measurements operate directly on individual functions and do not impose compilation constraints. Using Slither would have restricted the analysis to only the 198 compilable instances and would have provided contract-level rather than function-level metrics. However, to validate our implementation of Cyclomatic Complexity, we ran Slither on these 198 compilable cases and confirmed that the scores were equivalent.

To answer RQ$_4$, we report descriptive statistics for both complexity metrics across:  
(i) all 500 generated functions,  
(ii) the subset of functions that were deemed functionally correct according to RQ$_2$, and  
(iii) the ground-truth reference functions.  
We also compute both metrics for the functions generated using the RAG strategy to assess whether retrieval-augmented generation affects code complexity. This enables a comprehensive comparison between baseline and RAG settings and between generated and real-world functions.

\section{Study Result}
We report below the study results. For each RQ, we report both the zero-shot and RAG-based results.
\begin{figure*}
  \centering
  \setlength{\tabcolsep}{2pt} 
  \begin{tabular}{lll}
    \includegraphics[width=0.3\linewidth]{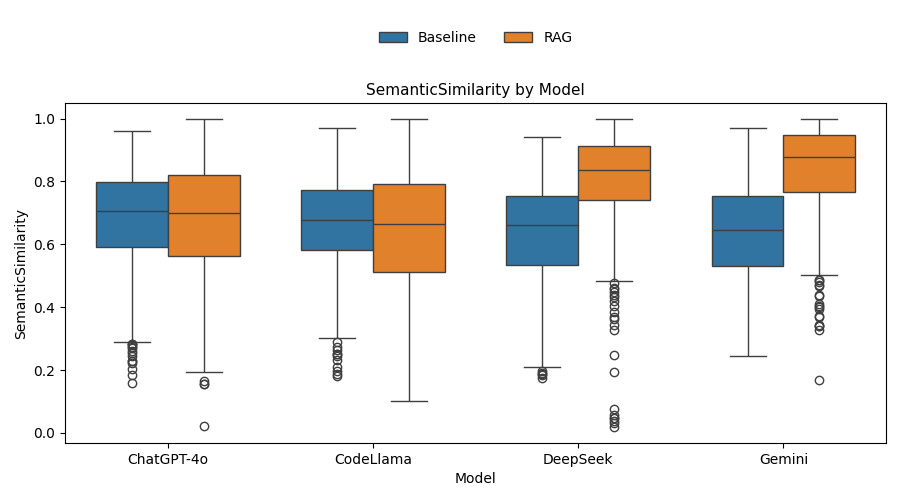} &
    \includegraphics[width=0.3\linewidth]{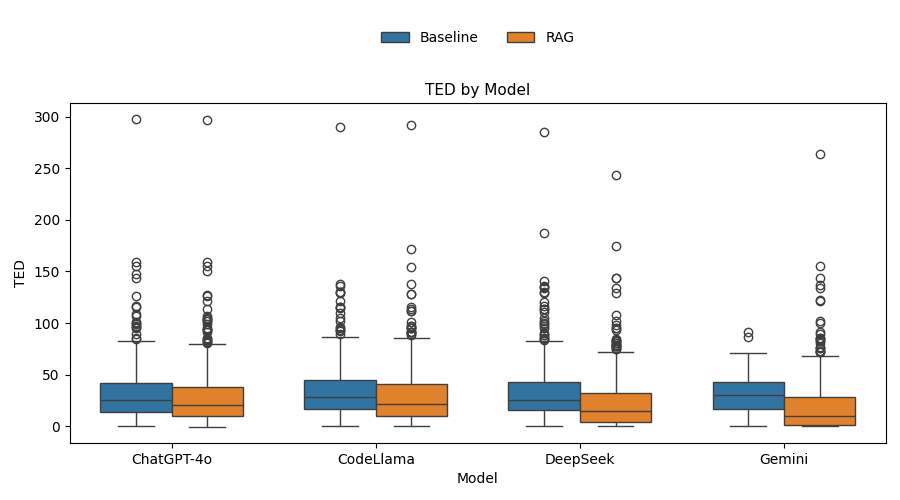} &
    \includegraphics[width=0.3\linewidth]{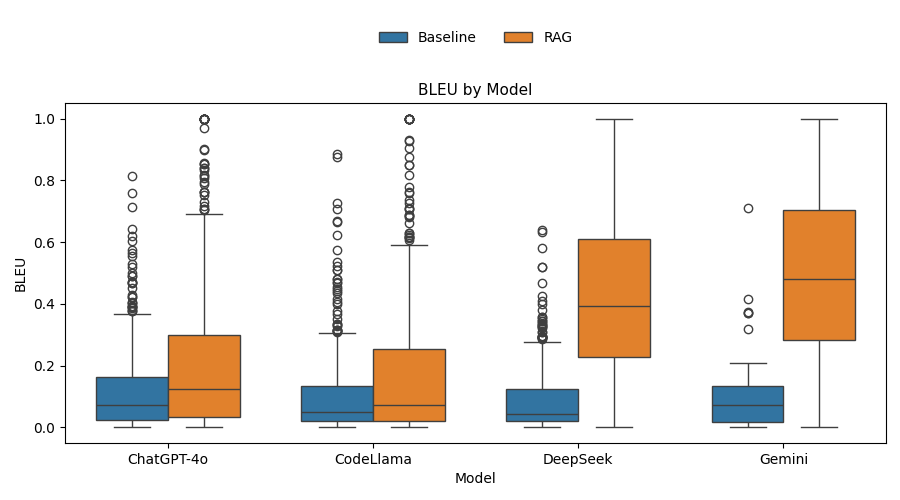} \\
    \small (a) Semantic Similarity distribution. &
    \small (b) TED metric distribution. &
    \small (c) BLEU score distribution. \\[2pt]

    \includegraphics[width=0.3\linewidth]{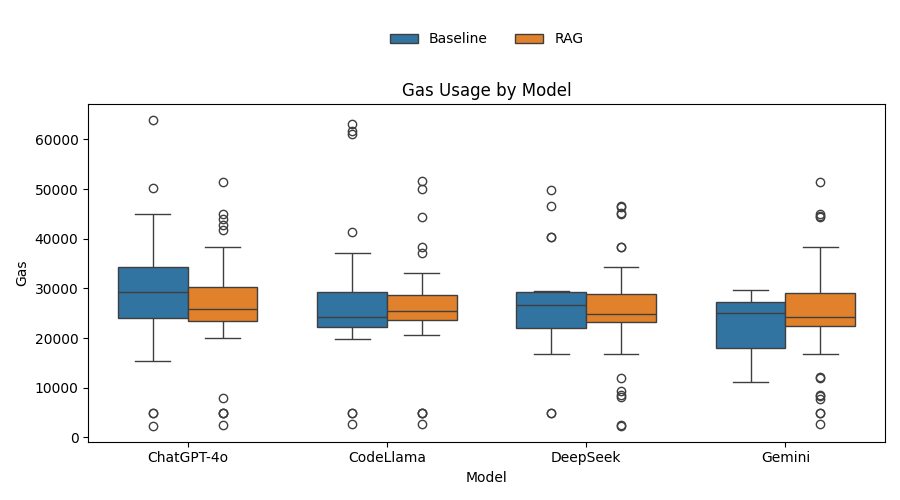} &
    \includegraphics[width=0.3\linewidth]{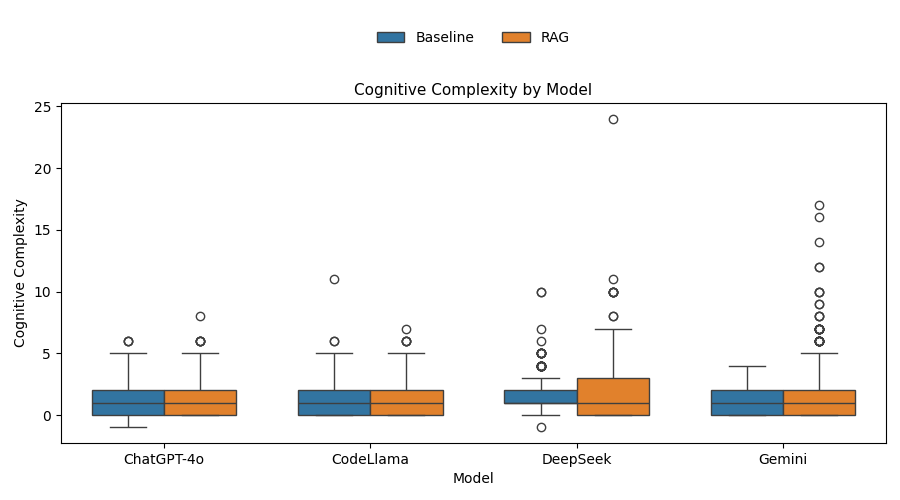} &
    \includegraphics[width=0.3\linewidth]{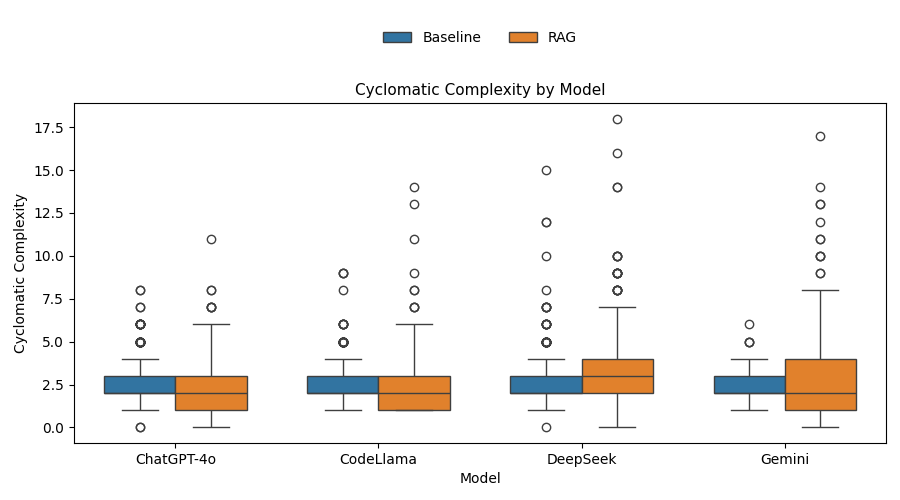} \\
    \small (d) Gas usage distribution. &
    \small (e) Cognitive complexity distribution. &
    \small (f) Cyclomatic complexity distribution.\\
  \end{tabular}
  \caption{Distributions of functional and complexity metrics across models and mode (Baseline vs RAG).}
  \label{fig:six_plots}
\end{figure*}

We also summarize all the results in Figure~\ref{fig:six_plots}, which reports boxplots of all the quantitative metrics involved in this study.

\subsection{\textbf{RQ$_1$}: Code Similarity}
Table~\ref{tab:similarity_metrics_horizontal} reports the similarity results for the code generated by models across semantic similarity, TED, and BLEU metrics.

\textbf{Zero-Shot.}
We start by discussing results regarding the baseline scenario. Among all evaluated models, ChatGPT-4o achieves the highest average semantic similarity with a mean of 0.6813, followed closely by Gemini (0.6707) and CodeLlama (0.6674). Deepseek-coder-v2 attains a slightly lower mean similarity of 0.6360. Overall, all models demonstrate the ability to generate code that is substantially semantically aligned with the ground truth, with mean scores consistently exceeding 0.6. The minimum similarity values span a broader range, from as low as 0.1584 (ChatGPT-4o) to 0.1811 (CodeLlama), highlighting that, in some instances, the generated functions diverge considerably from the intended semantics. By contrast, all models achieve maximum similarity values above 0.94, confirming that near-perfect semantic matches are possible across all approaches. The standard deviations, ranging between 0.1525 and 0.1642, indicate a moderate variability in semantic fidelity, likely influenced by differences in prompt complexity and the diversity of contract structures. Overall, these results confirm that, despite occasional discrepancies, the models generally succeed in capturing the semantic intent of the original code.

In terms of structural similarity, the TED results are also reported in Table~\ref{tab:similarity_metrics_horizontal}. The lowest average TED score is observed for ChatGPT-4o with a mean of 31.6680, indicating that, on average, it produces code that is structurally closer to the reference contracts. The other models follow with very similar mean scores, ranging from 33.3541 (Deepseek-coder-v2) to 34.0905 (CodeLlama). The relatively small differences in average TED suggest a comparable level of structural fidelity across models. All models exhibit a minimum TED of 0.0, reflecting perfect structural alignment in at least one example. Conversely, the maximum values vary between 279.0 (Gemini) and 298.0 (ChatGPT-4o), indicating that, in certain cases, the structural divergence from the ground truth can be substantial. Standard deviations are consistently high (approximately 27–28), suggesting a substantial spread of structural similarity scores across instances. These results indicate that, while average structural alignment is moderate, individual generations can display significant structural variance.

Finally, Table~\ref{tab:similarity_metrics_horizontal} also presents BLEU scores, which measure token-level n-gram overlap with the reference code. The highest average BLEU score is observed for ChatGPT-4o (mean = 0.1202), followed by Gemini (0.1087), CodeLlama (0.1055), and Deepseek-coder-v2 (0.0887). Although mean BLEU values are generally low, all models achieve high maximum BLEU scores, with CodeLlama reaching up to 0.8844, followed by Gemini (0.8555) and ChatGPT-4o (0.8154). These high values indicate that, in certain cases, the generated code is nearly identical to the reference in terms of token sequences. Interestingly, this indirectly means that no model can generate an exact match for any SC we considered. All models have a minimum BLEU of 0.0, confirming that some generations are completely disconnected from the SC to be generated. The standard deviation values (ranging from 0.1043 to 0.1343) indicate moderate variation in textual similarity across examples.

\textbf{RAG.}
The models with RAG tend to show higher semantic similarity and BLEU scores while achieving lower TED as compared to the zero-shot scenario, suggesting closer alignment with the ground truth both semantically and textually. For instance, Gemini with RAG achieves the highest average semantic similarity (\textbf{0.8392}) and BLEU (\textbf{0.4908}), while also reducing the mean TED to \textbf{19.1} from \textbf{33.7} in the base Gemini model. Similar trends are observed for Deepseek-coder-v2 with RAG, which raises its semantic similarity mean from \textbf{0.636} to \textbf{0.8006} and BLEU from \textbf{0.0887} to \textbf{0.4189}, reflecting substantially better alignment.
In contrast, ChatGPT-4o shows more modest improvements (semantic similarity increasing from \textbf{0.6813} to \textbf{0.6859}) and relatively stable BLEU and TED scores, suggesting less sensitivity to added comments in its generation process.
Notably, all improved models now reach a maximum semantic similarity of 1.0, indicating that the enrichment strategy enables near-perfect matches in some cases. However, standard deviations also tend to increase (especially in BLEU), highlighting that improvements are not uniform across all examples.

Overall, the results suggest that, in both scenarios, while the models consistently produce semantically and structurally similar code on average, substantial variation remains across individual generations. Moreover, our findings demonstrate that augmenting prompts with richer comments can consistently boost the semantic and textual fidelity of generated code, even though the impact varies by model.

\begin{table}
\caption{Detailed similarity statistics for considered models across three metrics: Semantic Similarity, TED, and BLEU.}
\label{tab:similarity_metrics_horizontal}
\centering
{\footnotesize
\begin{tabularx}{\textwidth}{X S[table-format=1.4] r r S[table-format=1.4] S[table-format=1.4] r r S[table-format=1.4] S[table-format=1.4] r r S[table-format=1.4]}
\toprule
\multirow{2}{*}{\textbf{Model}} &
\multicolumn{4}{c}{\textbf{Semantic Similarity}} &
\multicolumn{4}{c}{\textbf{TED}} &
\multicolumn{4}{c}{\textbf{BLEU}} \\
\cmidrule(rr){2-5} \cmidrule(rr){6-9} \cmidrule(rr){10-13}
& {Mean} & {Min} & {Max} & {StdDev}
& {Mean} & {Min} & {Max} & {StdDev}
& {Mean} & {Min} & {Max} & {StdDev} \\
\midrule
CodeLlama             & 0.6674 & 0.1811 & 0.9707 & 0.1525  & 34.0905 & 0.0   & 290.0 & 27.1880 & 0.1055 & 0.0   & 0.8844 & 0.1331 \\
CodeLlama-RAG       & 0.6434 & 0.1015 & 1.0000 & 0.2003  & 29.6076 & 0.0   & 292.0 & 28.7273 & 0.1743 & 0.0   & 1.0000 & 0.2283 \\
DeepSeek-Coder-v2     & 0.6360 & 0.1743 & 0.9420 & 0.1642  & 33.3541 & 0.0   & 285.0 & 27.9981 & 0.0887 & 0.0   & 0.6402 & 0.1043 \\
DeepSeek-Coder-v2-RAG & 0.8006 & 0.0176 & 1.0000 & 0.1699  & 22.7746 & 0.0   & 243.0 & 26.7864 & 0.4189 & 0.0   & 1.0000 & 0.2544 \\
Gemini                & 0.6707 & 0.1677 & 0.9813 & 0.1594  & 33.7505 & 0.0   & 279.0 & 28.2324 & 0.1087 & 0.0   & 0.8555 & 0.1322 \\
Gemini-RAG          & 0.8392 & 0.1685 & 1.0000 & 0.1438  & 19.1207 & 0.0   & 264.0 & 26.2697 & 0.4908 & 0.0   & 1.0000 & 0.2730 \\
ChatGPT-4o            & 0.6813 & 0.1584 & 0.9617 & 0.1562  & 31.6680 & 0.0   & 298.0 & 27.6579 & 0.1202 & 0.0   & 0.8154 & 0.1343 \\
ChatGPT-4o-RAG      & 0.6859 & 0.0235 & 1.0000 & 0.1869  & 28.2008 & 0.0   & 297.0 & 28.5822 & 0.2105 & 0.0   & 1.0000 & 0.2417 \\
\bottomrule
\end{tabularx}
}
\end{table}

\subsection{\textbf{RQ$_2$}: Code plausibility}
We report the number of contracts that are able to compile and be deployed, starting from the set of 198 compilable contracts from the ground truth, alongside the number and the percentage of output that reflects, for a given input, the same output as the ground truth functions.
Here, Table \ref{tab:correctness_combined} summarizes the functional plausibility of code generated by different LLMs.

\textbf{Zero-Shot.} 
In the baseline scenario, functional plausibility for models consistently shows low accuracy values. Gemini achieves the highest accuracy at 25.64\%, though it was evaluated on a notably smaller set of contracts (11) and samples (78) compared to other models. Notably, in this case, the number of compilable contracts decreases from 198 to only 11, which raises concerns about assessing generated code purely based on semantic and structural similarity.
CodeLlama and DeepSeek-Coder-v2exhibit similar performance, with 22.32\% and 21.71\% accuracy, respectively, from a larger pool of generated artifacts. ChatGPT shows the lowest accuracy at 20.79\%. These results indicate that generating functionally correct SCs from scratch remains a significant challenge for current LLMs, underscoring the need for augmented generation strategies. Generated instances with complete plausibility are considerably lower, ranging from about 7\% to 9\%. This discrepancy is largely due to subtle mismatches between the generated functions and the ground truth, especially in the way corner cases and input validation using \texttt{require} statements are managed. Providing insights into this, Figure~\ref{lst:gen_set_price}, which shows a ground truth function and a generated one, differs from the introduction of a \texttt{require} to validate input, which results in a diverse output when the input is $0$.

\begin{figure}[b]
\centering
\begin{lstlisting}[language=Solidity]
function setCurrentPrice(uint256 newPrice) public onlyOwner {
    currentPrice = newPrice;
}
\end{lstlisting}

\begin{lstlisting}[language=Solidity]
function setCurrentPrice(uint256 _currentPrice) public onlyOwner {
    require(_currentPrice > 0, "Price must be greater than zero");
    currentPrice = _currentPrice;
}
\end{lstlisting}
\caption{Ground-Truth and ChatGPT-generated function with input validation.}
\label{lst:gen_set_price}
\end{figure}

Notably, in this case, the LLM introduced a useful check, which was absent in the ground truth instance. However, in most cases, the LLM-generated version presents fewer validations than the ground truth corresponding one, as Figure~\ref{lst:gen_set_mtd} displays. 

\begin{figure}[b]
\centering
\begin{lstlisting}[language=Solidity]
function setMtdAmount(uint256 mtdAmountInWei) public onlyOwner {
    require(mtdAmountInWei > 0, "MTD amount must be greater than zero");
    require(mtdAmount != mtdAmountInWei, "MTD amount has not changed");
    mtdAmount = mtdAmountInWei;
    updatePrices();
}
\end{lstlisting}
\begin{lstlisting}[language=Solidity]
function setMtdAmount(uint256 _mtdAmount) public onlyOwner {
    require(_mtdAmount != 0, "Value must be greater than zero");
    mtdAmount = _mtdAmount;
}
\end{lstlisting}
\caption{Ground-Truth and ChatGPT-generated function.}
\label{lst:gen_set_mtd}
\end{figure}

\textbf{RAG.}
When applying RAG, we observe a substantial improvement in functional plausibility across all models. As reported in Table~\ref{tab:correctness_combined}, DeepSeek-Coder-v2-RAG demonstrates the most substantial improvement, more than doubling its accuracy from 21.71\% to 45.19\%. This significant gain suggests that DeepSeek-Coder-v2 particularly benefits from relevant code examples, translating into considerably more accurate functional behavior. Gemini-RAG also shows a notable increase, rising from 25.64\% to 37.01\%. Importantly, the number of generated and testable contracts for Gemini increased dramatically (from 11 to 130 contracts and 78 to 862 samples), indicating that RAG not only improves plausibility but also enables Gemini to produce a much larger volume of valid, testable code. ChatGPT-RAG improves from 20.79\% to 29.00\%, a substantial gain, accompanied by a significant expansion in the number of processed contracts and samples (from 59 to 101 contracts, and 606 to 938 samples). This suggests that RAG helps ChatGPT generate a broader range of robust code. CodeLlama-RAG exhibits a more modest increase, moving from 22.32\% to 23.56\%. While the accuracy improvement is slight, the marginal increase in processed samples (from 802 to 836) suggests a lesser, but still present, benefit from RAG in this context.

In conclusion, RAG integration proves to be a critical enabler for improving the functional plausibility of generated SCs. It provides essential context, enhancing models' understanding of intent and semantic fidelity. The results highlight varying degrees of benefit across models, with DeepSeek-Coder-v2 and Gemini showing the most pronounced gains.

\begin{table}
\centering
\caption{Plausibility results for all considered LLMs.}
\label{tab:correctness_combined}
\begin{tabular*}{\columnwidth}{@{\extracolsep{\fill}}lrrr}
\toprule
\textbf{Model} & \textbf{Contracts} & \textbf{Correct Calls (\%)} & \textbf{100\% Correct Contracts} \\
\midrule
CodeLlama       & 136 & 22.32\% & 7.35\% \\
CodeLlama-RAG   & 137 & 23.56\% & 9.49\% \\
DeepSeek-Coder-v2        & 140 & 21.71\% & 7.86\% \\
DeepSeek-Coder-v2-RAG    & 143 & 45.19\% & 23.08\% \\
ChatGPT         &  59 & 20.79\% & 8.47\% \\
ChatGPT-RAG     & 101 & 29.00\% & 19.80\% \\
Gemini          &  11 & 25.64\% & 9.09\% \\
Gemini-RAG      & 130 & 37.01\% & 13.85\% \\
\bottomrule
\end{tabular*}
\end{table}

\subsection{\textbf{RQ$_3$}: Gas Consumption }
In order to address RQ$_3$, we show the gas consumed by valid transactions on generated functions, alongside data regarding ground truth gas fees.
Hence, Table~\ref{tab:gas_combined} shows the gas consumption statistics for generated contracts. 

\textbf{Zero-Shot.}
In the baseline setting, we observe substantial variability across models, reflecting differences in generation strategies. The mean gas consumption is generally lower than that of the ground truth. Specifically, ChatGPT and CodeLlama exhibit similar average gas usage (approximately 28,000 units), while Gemini produces the most efficient code on average, with a mean consumption of around 21,900 units per invocation.

Minimum gas values range between approximately 2,200 and 11,000 units, while maximum consumption can reach nearly 64,000 units. These differences underscore the heterogeneous runtime costs produced by large language models. Notably, all models generate functions that, on average, consume less gas per call than the manually authored ground truth contracts. This result suggests that the generated implementations tend to produce simpler execution paths and potentially omit more sophisticated runtime checks or initialization logic, which may impact functionality or safety. The Mean High Consistency values report the average gas for functions matching ground truth outputs on at least 75\% of inputs. Notably, these means remain below the ground truth average, suggesting that even higher-fidelity generations tend to use less gas overall than manually authored contracts.

Compared to the non-RAG setting, all models show a consistent reduction in mean gas usage per invocation. For example, the average gas fees for ChatGPT decrease from approximately 28,900 to 26,600 units. Similar improvements are observed for CodeLlama (from ~27,800 to ~25,900) and DeepSeek-Coder-v2 (from ~26,700 to ~25,600).
These reductions suggest that retrieval augmentation enables models to generate more concise and efficient Solidity code. Notably, the gas consumption gap between models narrows when RAG is used, indicating a convergence toward simpler and more standardized outputs. Overall, these findings show that RAG generally contributes to modest gains in runtime efficiency, further lowering execution costs compared to baseline generation workflows without retrieval support. However, when focusing on the Mean High Consistency values, the improvements are less uniform. While ChatGPT and CodeLlama benefit from noticeable reductions, DeepSeek-Coder-v2 and Gemini show slightly higher values. This suggests that RAG may sometimes favor plausibility and completeness over raw gas optimization in functions that closely replicate reference behavior.

\begin{table}
\centering
\caption{Gas consumption statistics.}
\label{tab:gas_combined}
\begin{tabular*}{\columnwidth}{@{\extracolsep{\fill}}lrrrr}
\toprule
\textbf{Model} & \textbf{Min} & \textbf{Max} & \textbf{Mean} & \textbf{Mean High Cons.} \\
\midrule
Ground Truth        & 21,424.00  & 346,395.00 & 34,636.56 & -- \\
\midrule
ChatGPT             & 2,231.92   & 63,989.67  & 28,861.27 & 32,703.17 \\
ChatGPT-RAG         & 2,569.83   & 51,401.00  & 26,595.24 & 29,295.13 \\
CodeLlama           & 2,664.83   & 63,147.00  & 27,858.41 & 29,155.30 \\
CodeLlama-RAG       & 2,664.83   & 51,706.33  & 25,929.05 & 27,426.41 \\
Gemini              & 11,107.00  & 29,639.00  & 21,909.33 & 24,982.00 \\
Gemini-RAG          & 2,664.83   & 51,435.00  & 24,871.26 & 27,856.51 \\
DeepSeek-Coder-v2            & 4,875.83   & 49,897.50  & 26,734.44 & 27,155.22 \\
DeepSeek-Coder-v2-RAG        & 2,408.75   & 46,578.00  & 25,643.32 & 28,253.26 \\
\bottomrule
\end{tabular*}
\end{table}

\subsection{\textbf{RQ$_4$}: Complexity}
Answering \textbf{RQ$_4$}, we provide both cognitive and cyclomatic complexities for each model we used to provide insights into the complexity of generated functions.
To address this point, Table~\ref{tab:complexity_non_improved} reports the cognitive complexity statistics for the generated code, together with the distribution of differences compared to the ground truth. We start by discussing results about the baseline setting. Thus, all models produce functions that are less cognitively complex than the real contracts (mean = 3.00). Among them, CodeLlama exhibits the highest mean cognitive complexity (2.09), while ChatGPT-4o and Gemini produce simpler code with average values around 1.79.

The difference analysis shows that, on average, the generated code tends to be simpler. CodeLlama is the only model with a slightly positive mean difference (+0.18), suggesting that, in some cases, it can produce functions more complex than the originals. In contrast, Gemini and DeepSeek-Coder-v2 have more negative mean differences (-0.38 and -0.30, respectively), indicating a consistent simplification of the logic. The wide ranges observed between minimum and maximum differences highlight substantial variability: for example, DeepSeek-Coder-v2 generates code that can be up to 10 points simpler or 6 points more complex than the ground truth, depending on the instance.

Table~\ref{tab:complexity_non_improved} also shows the cyclomatic complexity results. All models again produce functions with simpler control flow compared to the ground truth (mean = 3.08). The average cyclomatic complexity of the generated functions ranges between 2.50 and 2.60, resulting in negative mean differences across all models. For example, DeepSeek-Coder-v2 is the model that comes closest to the ground truth in terms of branching logic, with a mean difference of -0.47 and a maximum observed cyclomatic complexity of 15. Conversely, Gemini and ChatGPT-4o tend to generate functions with consistently lower complexity, as indicated by the more negative mean differences (-0.57 and -0.56) and smaller maximum values.

When considering the Correct Only subset, which includes the functions where the corresponding ground truth contracts compiled successfully, the results are largely consistent with the overall trends but show slightly reduced complexity scores. For example, the mean cognitive complexity decreases across all models in this subset: CodeLlama drops from 2.09 to 1.98, DeepSeek-Coder-v2 from 1.91 to 1.88, and Gemini from 1.79 to 1.58. The same effect is observed in cyclomatic complexity, where the mean values also become marginally lower. These reductions suggest that generated code aligned with compilable reference implementations is, on average, simpler in both logic and control flow. This pattern likely reflects the relative simplicity of the subset of ground truth functions that compiled successfully, reinforcing the observation that generation outputs tend to emphasize straightforward structures over more complex patterns.

These findings indicate that the evaluated LLMs tend to generate implementations that are less complex both cognitively and structurally. This simpler style may improve readability and reduce the likelihood of introducing errors, but it could also limit their applicability in scenarios requiring sophisticated control flows or advanced design patterns.

On the other hand, introducing additional context through RAG resulted in higher variability and, in several cases, increased complexity. Gemini with RAG achieved a mean cognitive complexity (2.91) close to the ground truth (3.00), suggesting that richer prompts can encourage models to produce more complex logic. DeepSeek-Coder-v2 with RAG reached the highest maximum cognitive complexity (24), and it was the only model with a slightly positive mean difference (+0.10), indicating that some generated functions were more complex than the reference implementations. Conversely, ChatGPT-4o with RAG remained simpler on average (1.87) and showed a negative mean difference (-0.27).

In terms of cyclomatic complexity, DeepSeek-Coder-v2 with RAG also stands out as the only model exceeding the ground truth mean (+0.09). All other models continued to generate functions with simpler control flow, displaying negative mean differences between -0.16 and -0.77. The maximum cyclomatic complexity values increased substantially for all models, confirming that RAG can lead to richer and more diverse control structures. These results indicate that retrieval-augmented prompts have the potential to narrow the gap between generated and real-world code complexity, although the impact varies across models: while DeepSeek-Coder-v2 and Gemini show clear responsiveness to enriched input, ChatGPT-4o remains comparatively conservative. When considering the Correct Only subset, which includes only those generations associated with ground truth functions that compiled successfully, the results offer additional perspective. In this scenario, the models generally produce slightly lower cognitive and cyclomatic complexity scores compared to their full distributions. For example, Gemini shows a reduction in mean cognitive complexity from 2.91 to 2.39 when restricting analysis to this subset, while DeepSeek-Coder-v2 decreases from 2.64 to 2.36. Similarly, ChatGPT-4o shifts from 1.87 to 1.80. These reductions indicate that the functions linked to compilable ground truth implementations tend to be simpler on average. The same pattern is visible for cyclomatic complexity, where the mean values across all models decrease when measured only on this subset. 

\begin{table*}
\footnotesize
\centering
\caption{Cognitive and cyclomatic complexity statistics and differences compared to the ground truth. Values for "Correct Only" represent metrics calculated solely on compilable ground truth samples and related generations.}
\label{tab:complexity_non_improved}
\begin{tabular*}{\textwidth}{@{\extracolsep{\fill}}lrrrrrrrrrrrr}
\toprule
\multirow{2}{*}{\textbf{Model}} 
& \multicolumn{4}{c}{\textbf{Cognitive Complexity}} 
& \multicolumn{4}{c}{\textbf{Diff Cognitive}} 
& \multicolumn{2}{c}{\textbf{Cyclomatic Complexity}} 
& \multicolumn{2}{c}{\textbf{Diff Cyclomatic}} \\
\cmidrule(lr){2-5} \cmidrule(lr){6-9} \cmidrule(lr){10-11} \cmidrule(lr){12-13}
& \multicolumn{1}{c}{Mean} & \multicolumn{1}{c}{Min} & \multicolumn{1}{c}{Max} & \multicolumn{1}{c}{StdDev}
& \multicolumn{1}{c}{Mean} & \multicolumn{1}{c}{Min} & \multicolumn{1}{c}{Max} & \multicolumn{1}{c}{StdDev}
& \multicolumn{1}{c}{Mean} & \multicolumn{1}{c}{Max}
& \multicolumn{1}{c}{Mean} & \multicolumn{1}{c}{Max} \\
\midrule
CodeLlama     
& 2.09 & 1 & 11 & 1.20
& +0.09 & -8.0 & +6.0 & 2.36
& 2.56 & 9
& -0.52 & +8 \\
\quad \textit{Correct Only}
& \textit{1.98} & \textit{1} & \textit{5} & \textit{1.07}
& \textit{-1.02} & \textit{-9.0} & \textit{+4.0} & \textit{2.01}
& \textit{2.41} & \textit{9}
& \textit{-0.66} & \textit{+8} \\
CodeLlama RAG     
& 1.79 & 1 & 7 & 1.12
& -0.40 & -9.0 & +3.0 & 1.81
& 2.31 & 14
& -0.77 & +13 \\
\quad \textit{Correct Only}
& \textit{1.61} & \textit{1} & \textit{7} & \textit{0.91}
& \textit{-1.39} & \textit{-28.0} & \textit{+4.0} & \textit{2.00}
& \textit{2.12} & \textit{11}
& \textit{-0.96} & \textit{+10} \\
DeepSeek-Coder-v2 
& 1.91 & 1 & 10 & 1.23
& -1.09 & -10.0 & +6.0 & 2.77
& 2.60 & 15
& -0.48 & +14 \\
\quad \textit{Correct Only}
& \textit{1.88} & \textit{1} & \textit{6} & \textit{1.10}
& \textit{-1.12} & \textit{-11.0} & \textit{+5.0} & \textit{2.25}
& \textit{2.58} & \textit{12}
& \textit{-0.50} & \textit{+9} \\
DeepSeek-Coder-v2 RAG 
& 2.64 & 1 & 24 & 2.26
& +0.10 & -13.0 & +16.0 & 1.99
& 3.17 & 18
& +0.09 & +17 \\
\quad \textit{Correct Only}
& \textit{2.36} & \textit{1} & \textit{10} & \textit{1.75}
& \textit{-0.64} & \textit{-28.0} & \textit{+7.0} & \textit{1.77}
& \textit{2.68} & \textit{10}
& \textit{-0.39} & \textit{+9} \\
Gemini        
& 1.79 & 1 & 4 & 0.79
& -1.21 & -4.0 & +3.0 & 2.13
& 2.51 & 6
& -0.57 & +5 \\
\quad \textit{Correct Only}
& \textit{1.58} & \textit{1} & \textit{3} & \textit{0.76}
& \textit{-1.41} & \textit{-5.0} & \textit{+2.0} & \textit{1.84}
& \textit{2.40} & \textit{5}
& \textit{-0.68} & \textit{+4} \\
Gemini RAG        
& 2.91 & 1 & 17 & 2.47
& -0.11 & -14.0 & +7.0 & 1.73
& 2.91 & 17
& -0.16 & +16 \\
\quad \textit{Correct Only}
& \textit{2.39} & \textit{1} & \textit{10} & \textit{2.00}
& \textit{-0.61} & \textit{-28.0} & \textit{+7.0} & \textit{1.76}
& \textit{2.36} & \textit{10}
& \textit{-0.72} & \textit{+9} \\
ChatGPT-4o    
& 1.79 & 1 & 6 & 1.04
& -1.21 & -7.0 & +4.0 & 1.98
& 2.51 & 8
& -0.57 & +7 \\
\quad \textit{Correct Only}
& \textit{1.76} & \textit{1} & \textit{6} & \textit{1.15}
& \textit{-1.24} & \textit{-8.0} & \textit{+3.0} & \textit{1.76}
& \textit{2.45} & \textit{8}
& \textit{-0.63} & \textit{+5} \\
ChatGPT-4o RAG    
& 1.87 & 1 & 8 & 1.14
& -0.27 & -8.0 & +6.0 & 1.67
& 2.40 & 11
& -0.67 & +10 \\
\quad \textit{Correct Only}
& \textit{1.80} & \textit{1} & \textit{6} & \textit{1.13}
& \textit{-1.20} & \textit{-28.0} & \textit{+3.0} & \textit{1.75}
& \textit{2.29} & \textit{11}
& \textit{-0.79} & \textit{+10} \\
\midrule
Ground Truth  
& 3.00 & 1 & 29 & 2.90
& -- & -- & -- & --
& 3.08 & 18
& -- & -- \\
\bottomrule
\end{tabular*}
\end{table*}

\section{Discussion}
We studied how several LLMs in different settings (zero-shot and RAG) are capable of generating SCs. Our results clearly show that, even when models produced Solidity functions with high semantic similarity scores compared to the ground truth, functional plausibility remained low (around 20–25\% without RAG).
This underscores the importance of evaluating functional outputs, as textual and structural similarity alone are insufficient proxies for functional equivalence~\cite{sharma2025assessing}. While ChatGPT-4o already showed high semantic similarity in the zero-shot setting, RAG led to only marginal gains in this metric. However, functional plausibility improved significantly with RAG, rising from 20\% to 29\%. The same happened for DeepSeek-Coder-v2, for which RAG almost doubled its plausibility from ~21\% to ~45\%. However, this improvement came with higher variability in complexity and occasional increases in cognitive complexity, suggesting that retrieval can both help models align better with real-world implementations.

Notably, DeepSeek-Coder-v2, which is fully open and self-hostable, achieved plausibility and complexity metrics competitive with or better than ChatGPT-4o when combined with RAG. This demonstrates that organizations can deploy high-quality generation pipelines using open models, narrowing the gap with proprietary solutions. In this sense, it is worth considering that such a model was ran on an Ubuntu 24 desktop, equipped with a Ryzen 9 and an NVIDIA RTX 4070, thus a consumer-level machine.
Generated functions tended to be simpler and more gas-efficient than ground truth implementations, likely because LLMs omit non-trivial checks and advanced patterns, as illustrated in Figure~\ref{lst:transfer2}. While this can reduce runtime costs, it raises concerns about the omission of critical security or plausibility safeguards.

In summary, we recommend practitioners to be very careful when using LLMs to generate SCs. While they are good at producing maintainable and reasonably gas-efficient solutions, they fall short in terms of plausibility and rarely produce directly-usable solutions. Thus, a proper inspection and check should be \textit{always} conducted before deploying automatically generated SCs.

\begin{figure}
\centering
\begin{lstlisting}[language=Solidity]
function transfer(address _to, uint256 _value) public returns (bool) {
    require(!funding_ended, "Funding has ended");
    require(msg.sender != founders, "Founders cannot transfer tokens");
    uint256 senderBalance = balances[msg.sender];
    require(senderBalance >= _value && _value > 0, "Insufficient balance or invalid value");
    balances[msg.sender] = safeSub(senderBalance, _value);
    balances[_to] = safeAdd(balances[_to],_value);
    emit Transfer(msg.sender, _to, _value);
    return true;
}
\end{lstlisting}

\begin{lstlisting}[language=Solidity]
function transfer(address _to, uint256 _value) public returns (bool) {
    require(_to != address(0), "Receiver address cannot be 0");
    require(balanceOf(msg.sender) >= _value, "Insufficient balance");
    _transfer(msg.sender, _to, _value);
    return true;
}
\end{lstlisting}
\caption{Example of Ground-Truth and ChatGPT-4o generated instance.}
\label{lst:transfer2}
\end{figure}

\section{Threats to Validity}
\textbf{Construct Validity.}  
Ground truth contracts were extracted and cleaned from an established dataset introduced by Hu et al.~\cite{hu2021automating}. This dataset primarily reflects Solidity practices up to that time and may not fully cover more recent language features. This creates a risk that the benchmarked ground truth could be partially outdated. However, we used Solidity 0.8+ to compile all SCs, thus establishing compatibility with modern compiler constraints and semantics. 

\textbf{Internal Validity.} The original dataset required preprocessing to resolve compilation issues, such as spacing artifacts around the dot operator (e.g., \texttt{this . balance}). We employed an LLM to automatically correct these formatting inconsistencies; although this process may introduce errors, manual inspection of the first 50 repaired functions confirmed the absence of impactful modifications.  

Semantic Similarity was computed using SmartEmbed~\cite{gao2019smartembed}, a model released in 2019 when Solidity syntax and idioms were less mature. This may introduce discrepancies between actual and measured similarity. To assess the magnitude of this issue, we manually examined 50 function pairs using three embedding models—CodeBERT, OpenAI \texttt{text-embedding-small}, and SmartEmbed. SmartEmbed proved the most reliable overall, whereas CodeBERT frequently assigned high similarity scores (>0.9) even to semantically divergent functions.  

Gas consumption was evaluated using Ganache, which does not perfectly reflect \textit{mainnet} execution. Nonetheless, Ganache is widely used as a reference testing environment~\cite{di2022profiling}. To reduce nondeterministic variability, each function was executed multiple times and average gas values were used. We reported gas results for both the full set of generated contracts and a subset of \textit{mostly correct} contracts (pass rate > 75\%). We did not restrict the analysis to fully correct contracts, as developers often consider trade-offs between robustness and efficiency (e.g., a function may fail in corner cases yet still offer more efficient execution paths).  

A potential threat to our evaluation arises from semantically correct yet non-identical code, such as generated contracts that introduce additional validation checks or slightly different control-flow patterns. To mitigate this, our analysis incorporates metrics that capture semantic rather than purely syntactic similarity (e.g., BLEU and Semantic Similarity), ensuring that minor deviations preserving intent are not excessively penalized. To further assess the impact of such variations, we compared the number of \texttt{require} statements—used as a proxy for input validation—between the original and generated functions. A paired t-test revealed no statistically significant difference (p-value $>$ 0.05) for 6 out of 8 model–setup combinations. For the remaining two (CodeLlama and DeepSeek-Coder-v2-RAG), the differences were statistically significant but exhibited negligible (-0.11) and small (0.26) effect sizes. These results indicate that deviations in validation logic are limited and do not materially affect the reliability of our correctness metrics.

\textbf{External Validity.} The findings of this study may not fully generalize to all blockchain ecosystems or SC programming languages. Our evaluation focuses exclusively on Solidity, which remains the dominant language for smart contract development~\cite{wohrer2020domain}, and is widely used across multiple platforms (e.g., Ethereum, Polygon). Nevertheless, further investigations are needed to assess whether the observed trends extend to contracts written in alternative languages or executed within different blockchain environments. 

The sample of 500 functions used in our evaluation may also not capture all characteristics of real-world SCs. To mitigate this threat, we extracted a statistically significant random sample from the available dataset, ensuring a 5\% margin of error at a 95\% confidence level. This provides reasonable coverage while preserving representativeness.

Even though the dataset originates from an earlier period of Solidity's evolution, its suitability is not substantially affected by language updates. Our study focuses on function-level generation conditioned on semantic specifications—an aspect largely independent of the Solidity version. All LLMs were explicitly instructed to generate code targeting Solidity~0.8 (the latest major version), and we verified compilation under this version. Recent work also shows that modern LLMs can reliably migrate Solidity code across compiler versions~\cite{ye2025bridging}, further supporting the relevance of our findings for future releases. Additionally, our evaluation employs semantic metrics (BLEU and Semantic Similarity), which are inherently robust to syntactic changes introduced by version differences.

\section{Conclusion}
\label{sec:conclusion}

This paper presented an empirical evaluation of four LLMs for Solidity smart-contract generation, assessed through structural and semantic similarity, functional plausibility, gas consumption, and cognitive complexity. Our results show that while current models can reproduce structural patterns with reasonable fidelity, fully correct functional behavior remains difficult to guarantee. Retrieval-augmented generation improves plausibility and tends to produce more efficient and less complex functions, but subtle validation and control-flow details are still frequently underspecified.

These observations highlight the need for more systematic approaches to guiding and refining LLM outputs in smart-contract development. Recent work on \emph{second-stage heuristic synthesis} \cite{pareschi2025entangled} offers a potential conceptual foundation for such guidance. Although originally developed in a different domain, its general strategy of representing and combining heuristics could, in future work, support more structured prompt design or refinement loops in Solidity code generation.

Exploring this connection is beyond the scope of the present study, but it may help inform more reliable and context-aware development workflows in subsequent research.

\begin{acks}
This work is funded by PRIN Project \textit{Trust Machines for TrustlessNess (TruMaN): The Impact of Distributed Trust on the Configuration of Blockchain Ecosystems} (Identifier Code 2022F5CLN2– CUP H53D23002400006) financed by the Italian Ministry of University. This study was partially funded through the NPRR project METROFOOD\-IT.  METROFOOD\-IT has received funding from the European Union---NextGenerationEU, NPRR---Mission 4 ``Education and Research'' Component 2: from research to business, Investment 3.1: Fund for the realisation of an integrated system of research and innovation infrastructures---IR0000033 (D.M. Prot. n.120 del 21 June 2022).
\end{acks}

\bibliographystyle{unsrt}
  \bibliography{sample-base}
\end{document}